\documentclass[11pt]{article}

\usepackage[letterpaper, margin=1in]{geometry}
\usepackage{setspace}
\usepackage{graphicx}
\graphicspath{{./images/}}
\usepackage{booktabs}
\usepackage{multirow}
\usepackage[table]{xcolor}
\usepackage{colortbl}
\usepackage{tikz}
\usetikzlibrary{arrows.meta}
\usepackage{array}
\usepackage{threeparttable}

\usepackage{mathtools}
\usepackage{esint}
\usepackage{amsmath, amssymb}
\usepackage{siunitx}
\usepackage{stackengine}
\stackMath

\usepackage{enumitem}
\usepackage{wrapfig}
\usepackage{float}
\usepackage{listings}
\usepackage{xspace}
\usepackage{marginnote}
\usepackage{comment}

\usepackage{caption}
\usepackage{subcaption}

\usepackage{hyperref}
\hypersetup{
  colorlinks=true,
  linkcolor=blue,
  citecolor=blue,
  urlcolor=blue
}
\usepackage{url}

\usepackage{pgfplots}
\pgfplotsset{compat=1.9, every axis/.append style={font=\scriptsize}}

\usepackage{natbib}
\title{Improving S\&P 500 Volatility Forecasting through\\Regime-Switching Methods}

\author{%
  Ava C. Blake\thanks{Contact author: \texttt{acb2309@columbia.edu}}%
  \and
  Nivika A. Gandhi\thanks{Contact author: \texttt{nag2175@columbia.edu}}%
  \and
  Anurag R. Jakkula\thanks{Contact author: \texttt{arj2161@columbia.edu}}%
}

\date{\today}

\begin{document}

\maketitle

\begin{abstract}
    Accurate prediction of financial market volatility is critical for risk management, derivatives pricing, and investment strategy. In this study, we propose a multitude of regime-switching methods to improve the prediction of S\&P 500 volatility by capturing structural changes in the market across time. We use eleven years of SPX data, from May 1st, 2014 to May 27th, 2025, to compute daily realized volatility (RV) from 5-minute intraday log returns, adjusted for irregular trading days. To enhance forecast accuracy, we engineered features to capture both historical dynamics and forward-looking market sentiment across regimes. The regime-switching methods include a soft Markov switching algorithm to estimate soft-regime probabilities; a distributional spectral clustering method that uses XGBoost to assign clusters at prediction time; and a coefficient-based soft regime algorithm that extracts HAR coefficients from time segments segmented through the Mood test and clusters through Bayesian GMM for soft regime weights and uses XGBoost to predict regime probabilities. Models were evaluated across three time periods—before, during, and after the COVID-19 pandemic. The coefficient-based clustering algorithm outperformed all other models, including the baseline autoregressive model, during all time periods. Additionally, each model was evaluated on its recursive forecasting performance for 5 and 10 day horizons during each time period. The findings of this study demonstrate the value of regime-aware modeling frameworks and soft clustering approaches in improving volatility forecasting, especially during periods of heightened uncertainty and structural change.
\end{abstract}

\newpage

\section{Introduction}

Traditional econometric models such as the Generalized Autoregressive Conditional Heteroskedasticity (GARCH) model and the Heterogeneous Autoregressive (HAR) model rely on strong assumptions of stationarity and homogeneity, which often fail in the presence of structural breaks, financial crises, or rapidly changing market conditions. The HAR model, introduced by \citet{corsi} in 2009, remains a widely used benchmark for modeling realized volatility using high-frequency data, relying on past volatility values over different time horizons (daily, weekly, monthly) to predict future volatility. The model takes the following form,

{\setlength{\abovedisplayskip}{8pt}
 \setlength{\belowdisplayskip}{8pt}
\begin{equation}
    RV_t = \beta_0 
    + \beta_d \cdot RV_{t-1} 
    + \beta_w \cdot \overline{RV}_{t-1}^{(w)} 
    + \beta_m \cdot \overline{RV}_{t-1}^{(m)} 
    + \varepsilon_t
\end{equation}
}

where $\beta_0, \beta_d, \beta_w, \beta_m$ are the HAR parameters to be estimated, and $\varepsilon_t$ is the error term\footnote{See Appendix~\ref{appx:features} for weekly and monthly Realized Volatility specifications.}. However, its fixed coefficient structure limits its ability to adapt to shifts between low and high volatility regimes, which are commonly observed in empirical financial data. Later findings suggest that this static dynamic structure does not fully capture the behavior of volatility, especially in regimes that switch between calm and turbulent periods.

Recent literature has begun to address this limitation by incorporating regime-switching dynamics into volatility forecasting models. For example, \citet{zhang_Chinese_International_regime_switching} implement a Markov regime-switching model in which HAR model parameters depend on unobservable market states. Their framework uses a two-regime Markov chain with statistically significant transition probabilities, showing that distinct regression coefficients between high- and low-volatility regimes improve model performance. The authors concluded that such regime-aware modeling outperforms the traditional HAR model, providing a strong justification for our exploration of regime-switching frameworks. However, Markov models often rely on hard regime assignments that ignore uncertainty in state transitions.

Our research builds on these findings by incorporating a more flexible soft-classification framework into the regime-switching architecture. Unlike \citet{zhang_Chinese_International_regime_switching}, we use posterior regime probabilities—soft weights—to construct forecasts that better reflect uncertainty in regime identification and transition dynamics. In our approach, each regime-specific model learns from the full dataset but emphasizes different segments according to probabilistic weights.

We further explore clustering as a form of regime-switching in our model architecture. Clustering aims to identify volatility regime labels for training and inference, raising key methodological questions: How is time-series data treated? How is the number of regimes determined? What clustering methods should be used? And how can clustering be validated beyond final regression accuracy? A notable challenge in this context is the breakdown of the i.i.d. assumption, since volatility data exhibit strong temporal dependencies. Nonetheless, clustering is employed precisely to capture these dependencies, as regimes emerge over time.

\citet{Prakash_2021} address this by using the Mood Test to segment time-series data based on variance shifts—a defining feature of regime transitions. They construct a distance matrix in Wasserstein space to quantify the effort required to morph one time-series segment into another, and apply spectral clustering to handle the nonlinear, sparse structure of financial data. Our second model architecture incorporates and advances these techniques by applying them directly to realized volatility forecasting. We extend their pipeline by training regime-specific HAR models on the clustered segments and using an XGBoost classifier to map current features to regime probabilities for forecasting.

Finally, we propose a third, novel model that addresses limitations of distributional-based clustering by clustering on regression coefficients instead. Specifically, we segment the time series, extract Ordinary Least Squares (OLS) regression coefficients for each segment, and cluster based on these coefficients. These vectors reflect how input features relate to future volatility, providing a more interpretable and dynamic basis for identifying regimes.

\section{Data Scope and Collection}
Our dataset comprises intraday price data in 5-minute intervals from the Standard \& Poor's 500 (or S\&P 500) Index. The 5-minute intervals were used in order to thoroughly capture intraday price dynamics and fluctuations. The data spans across the past 11 years from May 1st, 2014 to May 27th, 2025, accessed via the Bloomberg Terminal. Using \hbox{5-minute} closing prices, we calculate intraday log-returns as:

{\setlength{\abovedisplayskip}{8pt}
 \setlength{\belowdisplayskip}{8pt}
\begin{equation}
    r_{t,i} = \ln\left(\frac{P_{t,i}}{P_{t,i-1}}\right)
    \label{eq:returns}
\end{equation}
}

where $P_{t,i}$ is the price at the $i^\text{th}$ 5-minute interval on day $t$. Log returns were used as opposed to raw returns to ensure time-additivity and stabilize variance, which is a standard approach in financial modeling. These returns are then aggregated to compute daily realized volatility (RV), following calculation and normalization techniques widely accepted and used by \citet{zhang_Chinese_International_regime_switching}, \citet{luo_infinite_HAR_HMM}, \citet{ding_regine_switching}, \citet{li_HAR_LSTM}, \citet{hu_HAR_graphNN}, and many others. Additionally, RV values were adjusted to account for shorter trading days. 

{\setlength{\abovedisplayskip}{8pt}
 \setlength{\belowdisplayskip}{8pt}
\begin{equation}
    RV_t = \sqrt{\frac{N}{n} \sum_{i=1}^{n} r_{t,i}^2}
    \label{eq:RVt}
\end{equation}
}

Here, $n$ is the number of intraday returns observed on day $t$ (which may vary due to holidays), and $N$ is the standard number of returns in a full trading day (lasting from 09:30 EST until 16:00 EST)---78 for intraday data only, or 79 including overnight returns. This scaling ensures comparability of RV across days with varying lengths. The square root of realized variance was taken to obtain realized volatility, aligning the scale with standard deviation and improving interpretability in forecasting models. To ensure alignment across all variables, only days with available SPX data were retained in the dataset. Feature values were included only if they corresponded to these days. For any missing feature values on retained days, linear interpolation was used to impute the missing entries.

\section{Assumptions and Simplifications}
It is assumed that the 5-minute SPX closing prices are accurate and free of market microstructure noise, such as bid-ask bounce or discreteness of price changes. We also treat the closing price in each interval as a fair and consolidated measurement of market activity, ignoring any possible closing anomalies. Our calculation of daily real volatility as in Equation \eqref{eq:RVt} is assumed to capture the full variability of asset prices when returns are sampled at a high frequency. 

The HAR model extends a linear regression framework to forecast realized volatility and thus inherits the core assumptions of linear regression: homoskedasticity, implying constant variance of the residuals, that residuals are independent, and that residuals are normally distributed with a mean of zero. The HAR model further relies on the long- memory behavior of volatility, implying that past high volatility tends to predict present high volatility. In order to utilize a regime-switching Markov model, we assume the following: volatility follows different statistical behaviors (regimes) at different times and that the transitions between these regimes follow a Markov process - the probability of transitioning to the next regime depends only on the present regime. More broadly, we assume that these market regimes can be inferred from observable data. For financial time-series data, distribution shifts of the inputs and outputs often do occur, which is one defect of fixed-regime HAR models. Our paper addresses this issue by incorporating regime switching to account for one such distribution shift.

\section{Methodology}

The original HAR model was designed to capture realized volatility (RV) behavior across multiple time scales (daily, weekly, and monthly). These lags aim to address the idea that different types of investors operate on different time horizons, more accurately reflecting market dynamics. After building the HAR model, we assessed several statistics to evaluate the assumptions necessary for Ordinary Least Squares Regression. The Omnibus and Jarque-Bera tests strongly reject the null hypothesis of normally distributed residuals. The skewness and kurtosis values further reveal right-skewed and heavy-tailed residuals—features commonly observed in financial time series. Despite this, the Durbin-Watson statistic suggests no meaningful autocorrelation in the residuals. Additionally, a low condition number implies minimal multicollinearity among lagged volatility features. Overall, while the OLS HAR model captures some temporal structure, its failure to account for non-normality and fat tails highlights the need for regime-switching dynamics. Additionally, the ADF (Augmented Dickey-Fuller) test was performed on all input features to assess their stationarity. All features were found to be stationary.

\subsection{Feature Engineering}

We extend the HAR model by introducing a dual-memory structure that captures both historical volatility patterns and forward-looking market sentiment. Our model applies HAR-style lags to implied volatility features (VIX), allowing the model to respond to shifts in investor expectations. As implied volatility reflects future expectations of market volatility, it represents a significant forward-looking measure and can provide information not contained in historical past volatility. 

We introduce three additional features to capture nonlinear and asymmetric market behavior. We used the daily closing prices previously obtained to determined daily values for Jump Variation and Realized Kurtosis. 
Our engineered structure thus incorporates:
\begin{enumerate}
        \item Lagged VIX values (daily, 5-day, 22-day): market fear over varying time horizons
        \item Realized kurtosis: Measures tail risk and extreme return behavior
        \item Jump variation: Isolates large discontinuous price moves from continuous volatility
\end{enumerate}
This feature design allows the model to adapt across regimes by integrating both behavioral and structural signals, improving forecast accuracy under changing market conditions. \footnote{See Appendix~\ref{appx:features} for feature calculation details.}

            \begin{figure}[H]
                \includegraphics[width=1\linewidth]{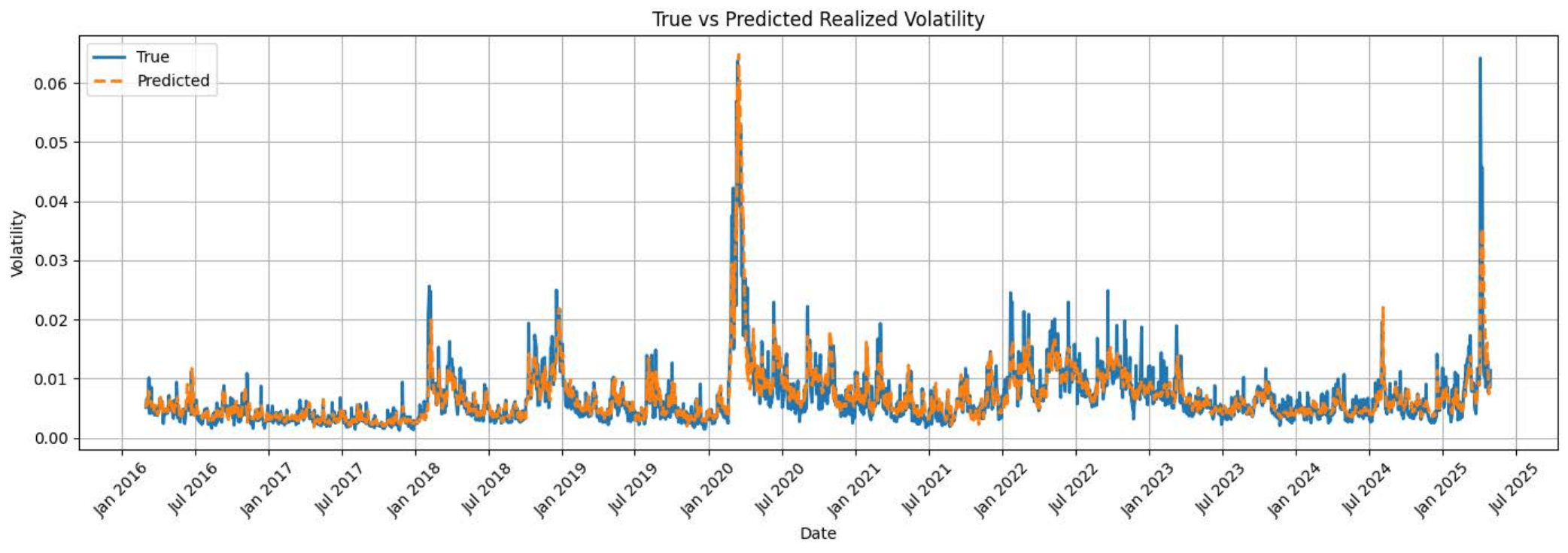} \\ 
                \caption{\small{Feature-engineered standard HAR model over the eleven-year period.}}
                \label{Figure 1}
            \end{figure}

We evaluated the standard HAR model against our feature-engineered model across the entire eleven-year period. Our findings establish that the feature-engineered model outperforms the standard HAR model in terms of Mean Squared Error (MSE) and Mean Absolute Percentage Error (MAPE). Figure \ref{Figure 1} illustrates the time series forecast from the feature-engineered HAR, demonstrating its robust ability to capture granular movements and some abrupt spikes. 

\begin{table}[H]
    \centering
    \small
    \begin{tabular}{l r r c}
        \toprule
        \textbf{Model} & \textbf{MAPE} & \textbf{MSE} \\
        \midrule
        HAR & 27.72 & 8.85 \\ 
        Feature-Engineered HAR & 26.37 & 8.35 \\
        \bottomrule
    \end{tabular}
    \caption{Models predict in 5-day increments. MSE values have been divided by 1,000,000 to improve numerical interpretability.}
\end{table}

Incorporating these additional features with a regime-switching model takes the following form: 
           
\begin{align}
\begin{split}
RV_t =\; & \beta_{0,S_t} 
+ \beta_{d,S_t} \cdot RV_{t-1} 
+ \beta_{w,S_t} \cdot \overline{RV}_{t-1}^{(w)} 
+ \beta_{m,S_t} \cdot \overline{RV}_{t-1}^{(m)} 
+ \gamma_{d,S_t} \cdot VIX_{t-1} \\
& + \gamma_{w,S_t} \cdot VIX_{t-1}^{(w)}
+ \gamma_{m,S_t} \cdot VIX_{t-1}^{(m)} 
+ \zeta_{S_t} \cdot KTS_{t-1}
+ \eta_{S_t} \cdot JMP_{t-1}
+ \varepsilon_t
\end{split}
\end{align}

where $VIX$ is the Chicago Board of Exchange's implied volatility, $KTS$ is realized kurtosis, and $JMP$ is the jump variation. The model coefficients  depend on the unobserved regime state $S_t$.

\subsection{Rolling Window Implementation}

For each model, we implement a rolling window framework to improve the model's adaptability to changing market conditions. Training on a fixed dataset can often lead to outdated parameter estimates when working with financial time series data. To address this limitation, we retrain our models on a fixed-length window of the past 441 observations (equivalent to 7 fiscal quarters) and generate forecasts for a shorter horizon of 5 or 10 days. We selected a window of 441 days to provide enough historical data for stable parameter estimation while still preserving responsiveness to regime changes. After generating these forecasts, we move the training window one step forward in time. Thus, the model continuously updates to reflect the most current market behavior. In particular, we employ a forecast horizon of 5 days for evaluation of our models during the Pre-COVID and Post-COVID periods, and a 10-day forecast horizon during the COVID period.

\subsection{Markov Regime-Switching Model}
We aim to capture how volatility behaves differently across changing market conditions by first developing a soft-regime-switching model. Instead of training a single global model, we fit a separate HAR model for each regime. A Hidden Markov Model (HMM) identifies these regimes and assigns soft probabilities to each time point, reflecting the likelihood of being in each regime. Our final forecast is therefore a weighted blend of the regime-specific HAR predictions, where the weights depend on the inferred regime probabilities.

The Hidden Markov Model (HMM) is a statistical model that represents a system with unobserved states, in this case, regimes, that evolve according to a Markov process. This framework looks to address the goal of accounting for latent structural shifts in market dynamics. Our algorithm aligns with previous work that incorporates the EM algorithm, such as \citet{zhang_Chinese_International_regime_switching}.

In each rolling training window, features $X$ and target RV array $y$ are z-score normalized, and a smoothed version of $y_t$, denoted $\tilde{y}_t$, is computed using a 5-day moving average. This computation occurs so that the HMM captures structural changes rather than noise. A Gaussian HMM with K regimes is fitted to $\tilde{y}_{1:t}$. Thus, the HMM models volatility as a sequence generated by an unobserved regime process, with each regime emitting Gaussian-distributed observations: 

\begin{align}
    \tilde{y}_t \mid z_t = k \sim \mathcal{N}(\mu_k, \sigma_k^2)
\end{align}

This HMM fitting process outputs two probabilistic components of the algorithm: posterior regime probabilities and transition matrix entries. The posterior regime probability at time t, defined by $\gamma_t(k) = P(z_t = k \mid \tilde{y}_{1:T})$, is a backward-looking, data-informed probability that describes how likely a regime $k$ is at time $t$ given all observed data up to the last time in the training window, $T$. The posterior regime probabilities provide the soft weights for fitting the regime-specific weighted least squares models at training time, as seen in the following minimization function: 

\begin{align}
    \min_{\{\beta^{(k)}\}} \sum_{t=1}^{T} \gamma_t(k) \left(y_t - x_t^\top \beta^{(k)}\right)^2
\end{align} 

If the data is sparse, the algorithm fallbacks to unweighted OLS. The transition matrix entries, with each entry expressed as $T_{ij} = P(z_t = j \mid z_{t-1} = i)$, gives the probability of transitioning from regime $i$ at time $t-1$ to regime $j$ at time $t$ at prediction time. For each step-ahead forecast h, the final prediction is a regime-weighted average of the HAR model outputs: 

\begin{align}
\hat{y}_{t+h} = \sum_{k=1}^{K} p_{t+h}(k) \cdot x_{t+h}^\top \beta^{(k)}.
\end{align}

The Markov switching mechanism is driven solely by the smoothed realized volatility series $\tilde{y}_t$, thus allowing regime inference to focus on structural shifts in volatility dynamics. However, the subsequent regime-specific HAR models are fitted using a larger feature set which includes lagged RV averages, lagged VIX averages, jump variation, and kurtosis. These features help to capture a broader range of market characteristics within each inferred regime.

\subsection{Distributional Clustering}
Unlike the Hidden Markov Model, which assumes that each regime follows a specific statistical distribution, our distributional clustering approach does not make assumptions about the underlying form of volatility behavior. Instead, this method identifies structural shifts in the data using a statistical test and then clusters based on the distribution of features. We fit a separate model for each identified regime and use a classifier to assign data points to the most likely regime.

The input features $X$ and target RVs $y$ are first z-score standardized. We apply the Mood's Median Test, a statistical test that detects distributional changes in variance, to segment the data. For each time t, we test for two-sample scale differences over a rolling window w:

\begin{align}
    H_0: \operatorname{Var}(\tilde{y}_{t-w:t}) = \operatorname{Var}(\tilde{y}_{t:t+w})\\
    H_1: \operatorname{Var}(\tilde{y}_{t-w:t}) \ne \operatorname{Var}(\tilde{y}_{t:t+w})
\end{align}

The time points where the null is rejected are labeled as change points. These change points divide the time series into segments. Each segment $s_i$ is represented by the joint distribution over features and target: $D_i = \{ (\tilde{x}_t, \tilde{y}_t) \mid t \in s_i \}$.
Then, we compute the pairwise squared 2-Wasserstein distances between time segments according to the following equation,

\begin{align}
    W_{ij}^2 = W_2^2(D_i, D_j) = \min_{\pi \in \Pi(a,b)} \sum \|u - v\|^2 \, \pi(u,v)
\end{align}

where $\pi$ is the optimal transport plan, and $a$, $b$ are uniform weights over the two segments. The Wasserstein distance represents a similarity metric between two probability distributions. The result of this computation is a symmetric distance matrix W. Then, we convert pairwise distances into pairwise similarities using the equation below:

\begin{align}
    K_{ij} = \exp\left(-\frac{W_{ij}^2}{2\sigma^2}\right)
\end{align}

We apply spectral clustering on the kernel matrix $\boldsymbol{K}$. This process produces segment-level regime labels from 1 through number of regimes $K$. These labels are then projected to the individual time-series points in each time segment.

For each regime $k$, we fit a separate OLS model for each cluster. We then train an XGBoost classifier on the rolling training window to map the feature values of the individual data points to the segment-level cluster labels\footnote{See Appendix~\ref{appx:hyperparams} for XGBoost hyperparameter details.}. This classifier is then used to determine hard cluster assignments at prediction time. The output is rescaled through the inverse of z-score normalization.

This model is designed to cluster on the daily, weekly, and monthly RV and VIX averages, however, the prediction step incorporates a larger feature set. In particular, the HAR forecasting models and the XGBoost classifier both include the jump variation, which helps capture sudden market shocks that may not be reflected in the smoothed volatility features used for clustering.

\subsection{Coefficient-Based Soft Clustering}
As opposed to our prior two methods, the coefficient-based soft clustering approach does not define regimes based on patterns in the observed data itself. Instead, this approach identifies regimes according to how the relationship between features and volatility evolves over time. We then segment the series based on variance shifts, extract regression coefficients from each segment, and cluster them to uncover distinct regimes. Forecasts are generated by weighting regime-specific predictions according to probabilities output by a trained classifier.

The input features $X$ and target RVs $y$ are z-score standardized. We apply the Mood's Median Test to detect statistically significant changes in variance and thereby identify segment boundaries. For each time t, we test for two-sample scale differences over a rolling window w: 

\begin{align}
    H_0: \operatorname{Var}(\tilde{y}_{t-w:t}) = \operatorname{Var}(\tilde{y}_{t:t+w})\\
    H_1: \operatorname{Var}(\tilde{y}_{t-w:t}) \ne \operatorname{Var}(\tilde{y}_{t:t+w})
\end{align}

The time points where the null is rejected are labeled as change points. These change points divide the time series into segments. Each segment $s_i$ is represented by the joint distribution over features and target: $D_i = \{ (\tilde{x}_t, \tilde{y}_t) \mid t \in s_i \}$.
For each segment $s_i$, we fit a linear OLS model:

\begin{align}
    y_t = \beta_0^{(i)} + \mathbf{x}_t^\top \boldsymbol{\beta}^{(i)} + \varepsilon_t, \quad \text{for } t \in s_i
\end{align}

The coefficient vectors serve as descriptors of local dynamics. If OLS fails due to rank deficiency, ridge regression is used as a fallback. Principal Component Analysis (PCA) is applied to reduce the dimensionality of the coefficient vectors before clustering. Next, we apply a Bayesian Gaussian Mixture Model to estimate posterior membership probabilities of each regime k: 

\begin{align}
    \gamma_i^{(k)} = P(z_i = k \mid \theta^{(i)}), \quad \text{with} \quad \sum_{k=1}^K \gamma_i^{(k)} = 1 
\end{align}

We fit a Weighted Least Squares model on the full training window according to:

\begin{align}
\min_{{\beta^{(k)}}} \sum_{t=1}^{T} \gamma_t(k) \left(y_t - x_t^\top \beta^{(k)}\right)^2
\end{align}

for each regime $k$. For out-of-sample prediction, we train an XGBoost classifier on the mean feature vectors of each segment to map inputs to cluster assignments\footnote{See Appendix~\ref{appx:hyperparams} for XGBoost hyperparameter details.}. At prediction time, the classifier receives the features of a time point as input and outputs a soft probability vector, which is used to weight regime-specific predictions:

\begin{align}
\hat{y}{t+h} = \sum{k=1}^{K} p_{t+h}(k) \cdot x_{t+h}^\top \beta^{(k)}.
\end{align}

The features used in this clustering method consist of lagged RV averages, lagged VIX averages, and realized kurtosis. These inputs capture both persistent volatility patterns and tail risk, enabling the segment-level models to differentiate between regimes with distinct higher-order dynamics.

\section{Results and Discussion}

Each regime-switching model was tested on three different time periods: pre-COVID, COVID, and post-COVID. The pre-COVID time period ranged from 2014-06-02 to 2018-05-21, COVID ranged from 2018-05-21 to 2020-09-29, and post-COVID ranged from 2020-09-29 to 2025-04-29. Date ranges were determined based on major shifts in market behavior and volatility patterns associated with the onset and recovery phases of the COVID-19 pandemic. Model performance was compared within each time period and evaluated based on Mean Squared Error (MSE) and Mean Absolute Percentage Error (MAPE). Predictions were made in 5-day increments for the pre-COVID and post-COVID time periods, and 10-day increments for the COVID time period in order to account for heightened volatility and reduced data stability during the crisis period, which warranted a slightly broader forecasting window.

\begin{table}[H]
    \centering
    \small
    \begin{tabular}{l r r c}
        \toprule
        \textbf{Model} & \textbf{MAPE} & \textbf{MSE} & \textbf{\# Reg.} \\
        \midrule
        HAR & 27.09 & 3.40 & 1 \\ 
        Markov Soft EM & 24.33 & 3.12 &  3\\
        Distributional Clustering & 25.89 & 3.55 &  2\\ \rowcolor{yellow}
        Coefficient Clustering & 23.92 & 3.11 & 2 \\
        \bottomrule
    \end{tabular}
    \caption{Models predict in 5-day increments. MSE values have been divided by 1,000,000 to improve numerical interpretability.}
\end{table}

\begin{table}[H]
    \centering
    \small
    \begin{tabular}{l r r c}
        \toprule
        \textbf{Model} & \textbf{MAPE} & \textbf{MSE} & \textbf{\# Reg.} \\
        \midrule
        HAR & 30.13 & 35.35 & 1 \\
        Markov Soft EM & 31.93 & 36.22 & 2 \\
        Distributional Clustering & 32.63 & 33.59 & 2 \\
        \rowcolor{yellow}
        Coefficient Clustering & 30.62 & 31.91 & 2 \\
        \bottomrule
    \end{tabular}
    \caption{Models predict in 10-day increments. MSE values have been divided by 1,000,000 to improve numerical interpretability.}
\end{table}

\begin{table}[H]
    \centering
    \small
    \begin{tabular}{l r r c}
        \toprule
        \textbf{Model} & \textbf{MAPE} & \textbf{MSE} & \textbf{\# Reg.} \\
        \midrule
        HAR & 23.31 & 8.63 & 1 \\
        Markov Soft EM & 22.45 & 7.66 & 2 \\
        Distributional Clustering & 23.30 & 7.91 & 2 \\
        \rowcolor{yellow}
        Coefficient Clustering & 22.70 & 7.60 & 2 \\
        \bottomrule
    \end{tabular}
    \caption{Models predict in 5-day increments. MSE values have been divided by 1,000,000 to improve numerical interpretability.}
\end{table}

\begin{figure*}[ht]
    \centering

    \begin{subfigure}[t]{0.48\textwidth}
        \centering
        \includegraphics[width=\linewidth]{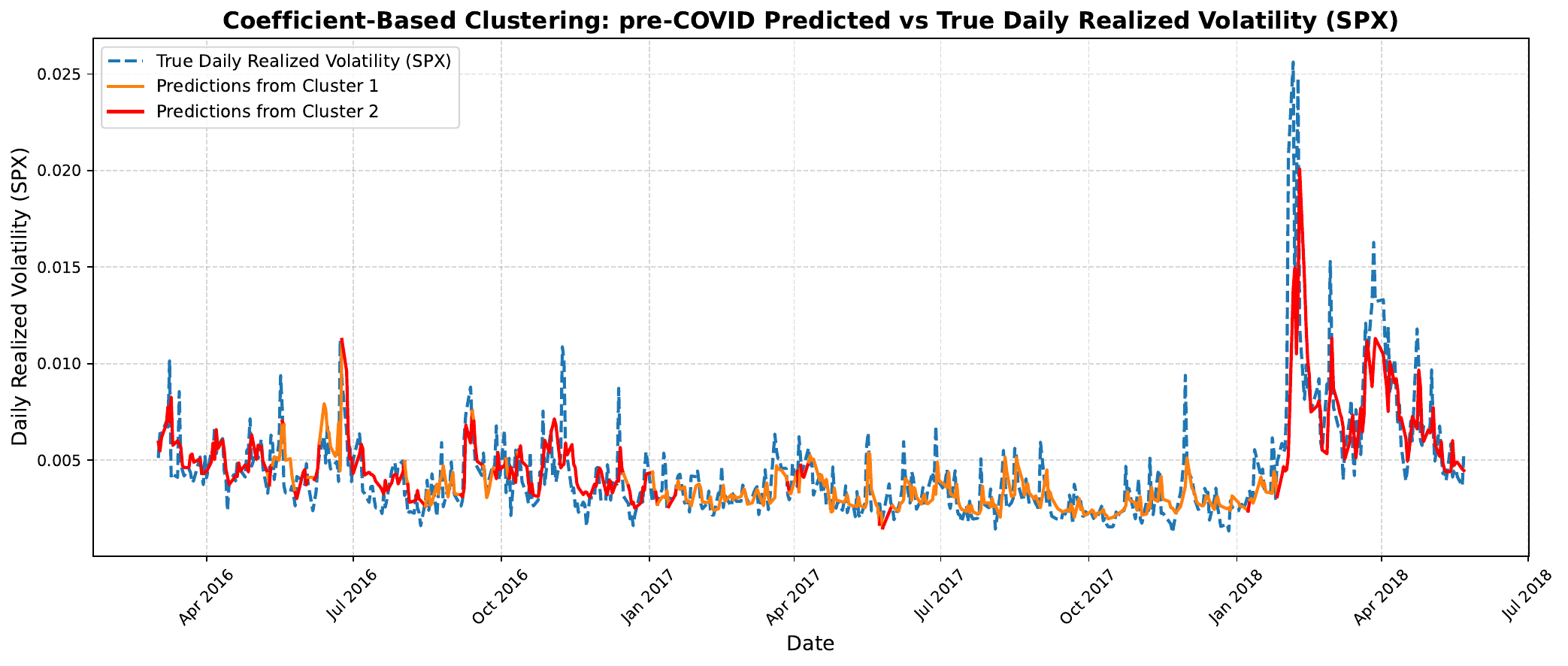}
        \caption{Pre-COVID: Coefficient-Based Clustering.}
        \label{fig:pre}
    \end{subfigure}%
    \hfill
    \begin{subfigure}[t]{0.48\textwidth}
        \centering
        \includegraphics[width=\linewidth]{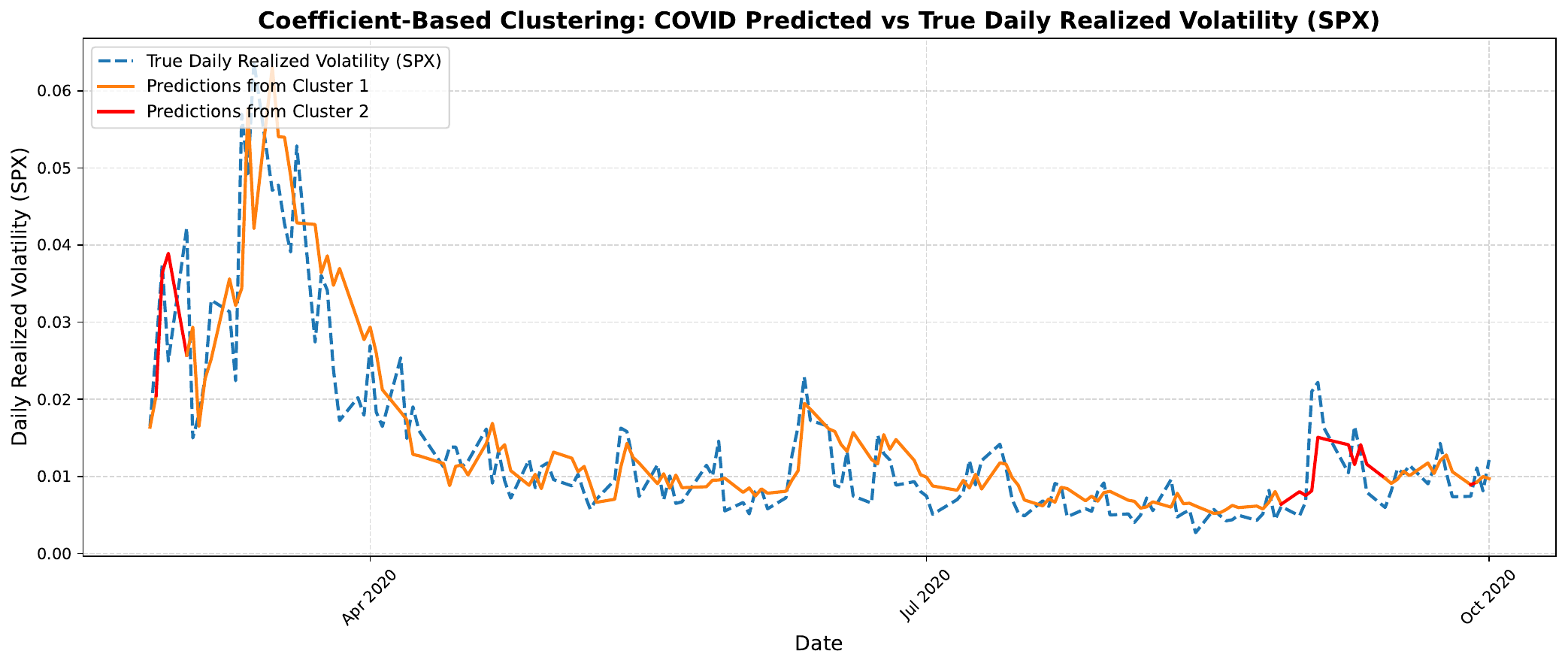}
        \caption{COVID: Coefficient-Based Clustering.}
        \label{fig:covid}
    \end{subfigure}

    \vspace{1em}

    \begin{subfigure}[t]{0.6\textwidth}
        \centering
        \includegraphics[width=\linewidth]{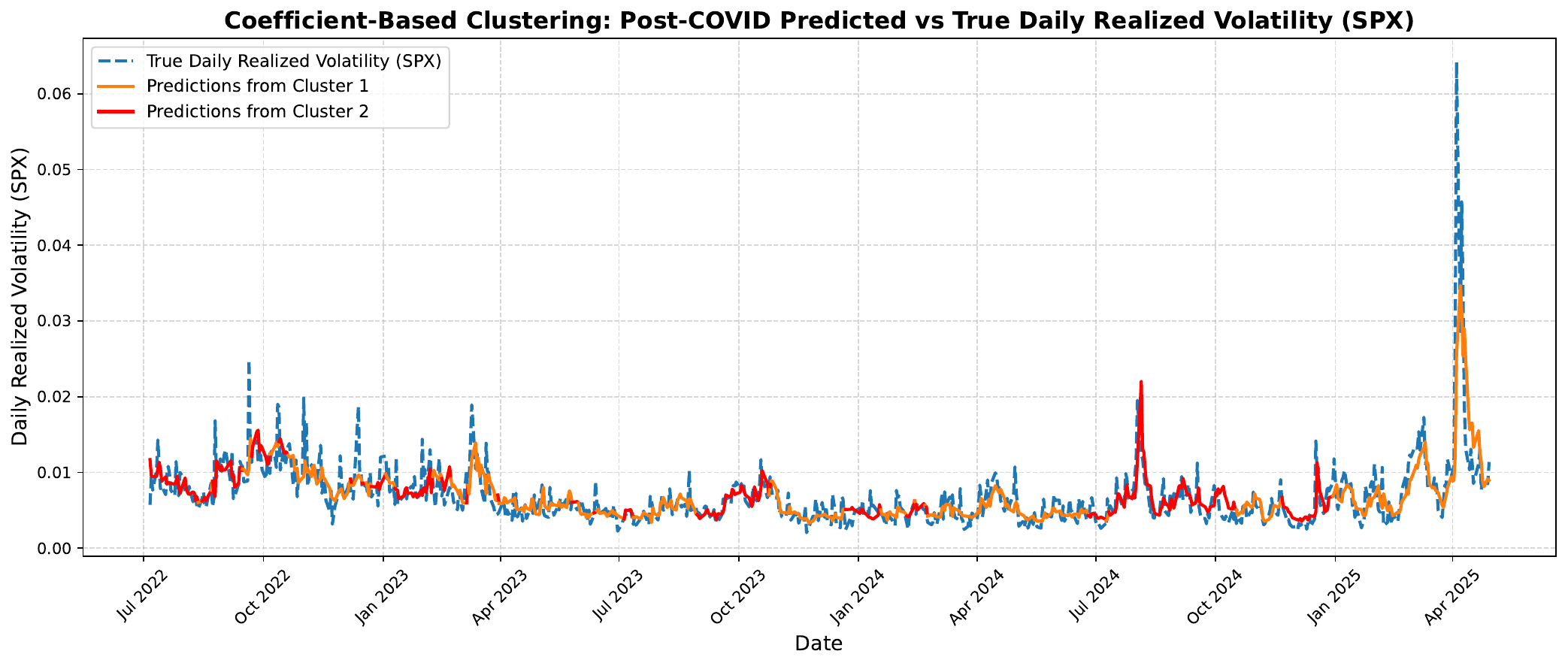}
        \caption{Post-COVID: Coefficient-Based Clustering.}
        \label{fig:post}
    \end{subfigure}

    \caption{Forecasts from the Coefficient-Based Clustering model across all periods.}
    \label{fig:clustering_all}
\end{figure*}

Across all periods, regime-switching models generally outperformed the baseline HAR model. In particular, the Coefficient-Based Clustering model achieved the lowest MSE during all time periods. It consistently reduced MSE compared to the standard HAR model, with improvements of 8.52\% (Pre-COVID), 9.73\% (COVID), and 11.9\% (Post-COVID). MAPE values showed a similar pattern in the pre- and post-COVID periods, confirming improved predictive accuracy for short-term volatility movements. The Markov regime-switching followed closely in the pre-COVID and post-COVID time periods, outperforming the standard non-regime-switching model in both time periods. The highly volatile COVID time period proved to be more difficult for the Markov HAR to predict, and the standard model was marginally more accurate. This may reflect the Markov model's limitations in adapting quickly to sharp, irregular market transitions during crisis periods. However, the Distributional Clustering model and Coefficient-Based Clustering model were better suited to handle abrupt structural breaks and regime shifts in volatility, leading to slightly improved performance during this unstable period. For most models, the optimal number of regimes was between two and three, suggesting that market dynamics can often be meaningfully described by a few distinct volatility states. Overall, the consistent reduction in MSE across all periods suggests that the regime-switching models not only improve average forecast accuracy but also better controls for large mispredictions, which is an important property to address for in volatility forecasting. We believe the Coefficient-Based Clustering method likely performed best because it captures the direct relationship between input features and volatility dynamics, leading to more informative regime distinctions. Figures \ref{fig:pre} through \ref{fig:post} illustrate the time series forecasts from the top-performing models in each period. Here, we can see how the Coefficient-Based Clustering model visibly adapts to local structural shifts across time periods. 

\subsection{Recursive Forecasting}

We extend the HAR framework by generating multi-step-ahead forecasts of realized volatility through a recursive procedure, where each forecasted value is fed into the model as an input for predicting the next time step. This allows the model to capture compounding effects and evolving volatility patterns beyond one-day horizons.

To more accurately capture the interdependent dynamics between realized and implied volatility, we develop a dual-recursive HAR-VIX framework. This approach jointly models and forecasts RV and VIX over time. At each recursive forecasting step, we first generate a forecast for VIX using both its own lags and the RV lags. This forecasted VIX is then used as an input, along with RV lags, to predict RV. The process repeats recursively, allowing the predictions for the two volatility measures to co-evolve and influence each other dynamically. This structure enables the model to capture feedback effects between market expectations (VIX) and realized outcomes (RV), which are often observed during periods of market stress or change. The only features in the recursive models are those which are being forecasted, as using feature's real values introduces look-ahead bias as the model forecasts future days based on previous predictions. The dual-recursive structure thus enables the use of VIX as a feature without introducing look-ahead bias, as it ensures that only forecasted values are used at each step while still allowing for features to enhance model accuracy.

Both the single and dual-recursive HAR frameworks are implemented within our regime-switching modeling pipelines. For each regime, determined via segmentation and clustering methods previously discussed, we estimate regime-specific model parameters. The dual-recursive approach is particularly well-suited to this regime-aware setting, as it accounts for dynamic interactions between RV and VIX within each structural regime.

\begin{table}[H]
  \centering
  \begin{tabular}{lccc|ccc|ccc}
    \toprule
    \multicolumn{1}{c}{Model} & \multicolumn{3}{c}{Pre-COVID} & \multicolumn{3}{c}{COVID} & \multicolumn{3}{c}{Post-COVID}\\
    \midrule
    Single-Recursive & MAPE & MSE & \# Reg. & MAPE & MSE & \# Reg. & MAPE & MSE & \# Reg. \\
    \midrule
    Standard HAR        & 33.90 & 5.10 & 1 & 35.68 & 67.06 & 1 & 31.22 & 16.97 & 1 \\
    Markov Soft EM      & 33.89 & 5.10 & 2 & 35.34 & 64.70 & 2 & 31.02 & 16.67 & 2 \\
    Dist. Clustering    & 34.97 & 5.46 & 2 & 32.47 & 62.49 & 2 & 32.38 & 19.62 & 2 \\
    Coeff. Clustering   & 34.97 & 4.61 & 2 & 35.23 & 65.34 & 2 & 30.63 & 13.57 & 2 \\
    \midrule
    Dual-Recursive & MAPE & MSE & \# Reg. & MAPE & MSE & \# Reg. & MAPE & MSE & \# Reg. \\
    \midrule
    Standard HAR        & 33.28 & 4.96 & 1 & 42.91 & 80.53 & 1 & 28.49 & 13.37 & 1 \\
    Markov Soft EM      & 32.33 & 4.91 & 2 & 54.48 & 88.01 & 2 & 27.55 & 12.80 & 2 \\
    Dist. Clustering    & 32.21 & 5.23 & 2 & 41.70 & \cellcolor{yellow}57.28 & 2 & 29.32 & 15.00 & 2 \\
    Coeff. Clustering   & 33.01 & \cellcolor{yellow}4.46 & 2 & 41.52 & 116.5 & 2 & 28.67 & \cellcolor{yellow}12.17 & 2 \\
    \bottomrule
  \end{tabular}
  \caption{Model performance for 5-day forecasts during each time period. Divide MSE values by 1,000,000. Performance is compared based on MSE values.}
  \label{tab:5-day-recursive}
\end{table}

\begin{figure*}[!t]
    \centering

    \begin{subfigure}[t]{0.48\textwidth}
        \centering
        \includegraphics[width=\linewidth]{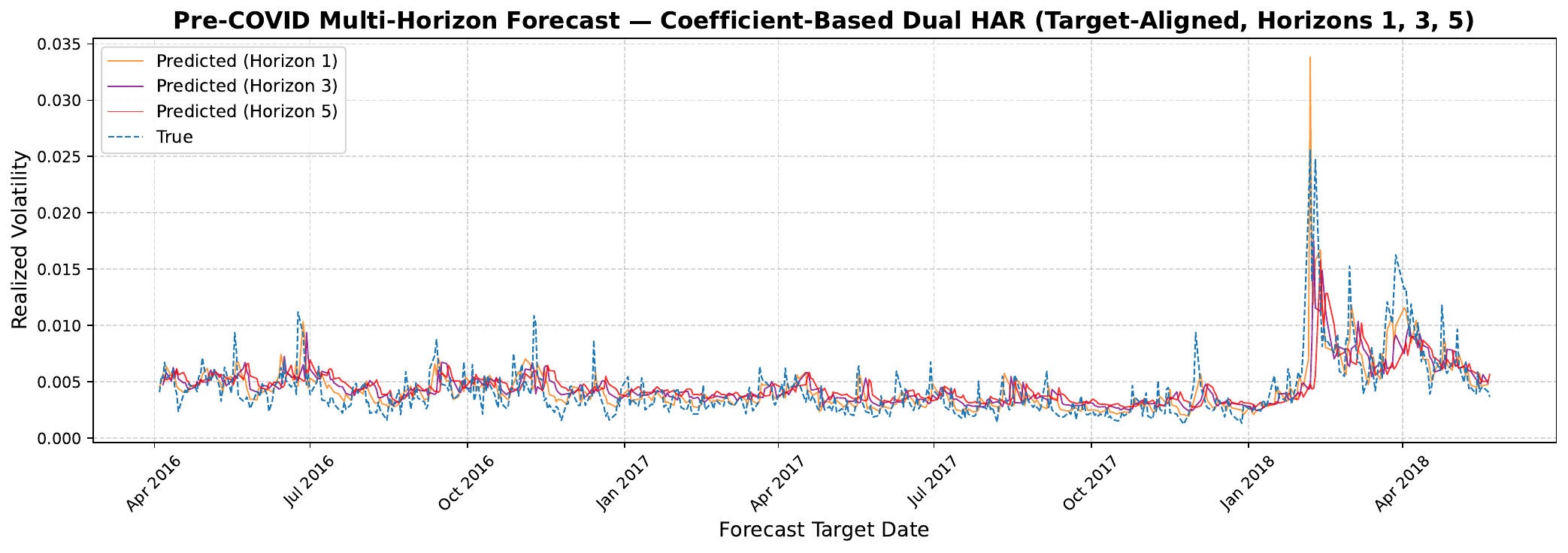}
        \caption{Pre-COVID: Dual-Recursive Coefficient-Based Clustering.}
        \label{fig:pre-5day}
    \end{subfigure}%
    \hfill
    \begin{subfigure}[t]{0.48\textwidth}
        \centering
        \includegraphics[width=\linewidth]{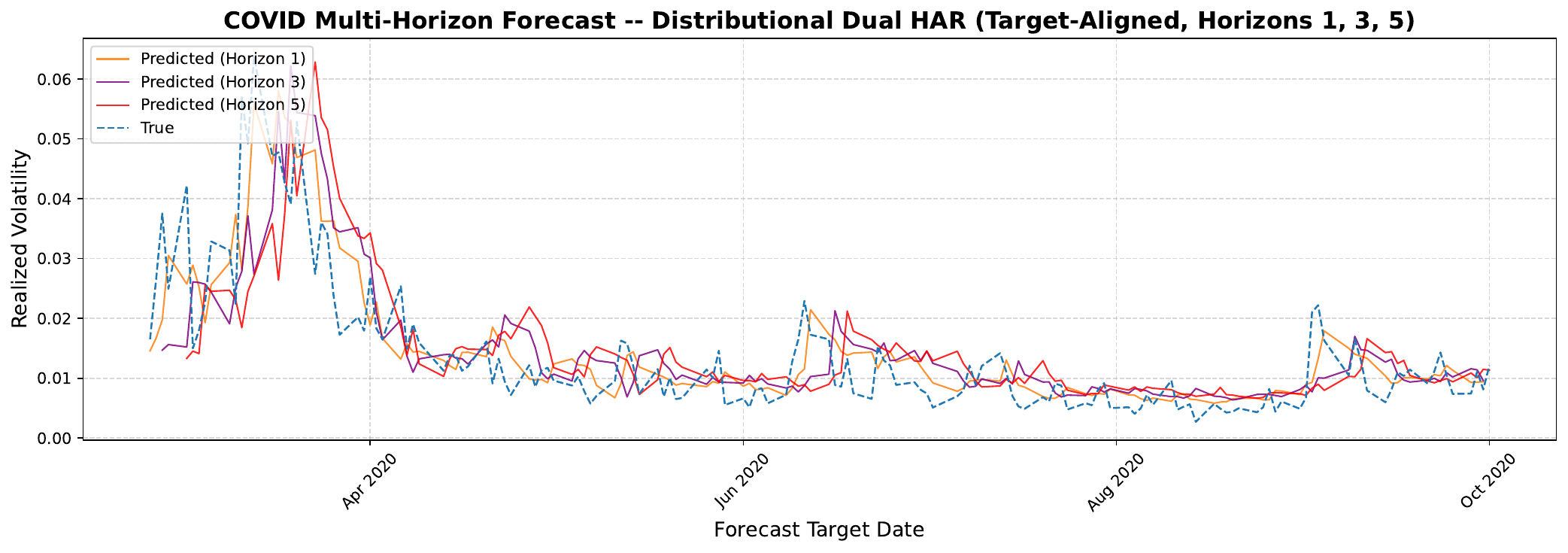}
        \caption{COVID: Dual-Recursive Distributional Clustering.}
        \label{fig:covid-5day}
    \end{subfigure}

    \vspace{1em}

    \begin{subfigure}[t]{0.6\textwidth}
        \centering
        \includegraphics[width=\linewidth]{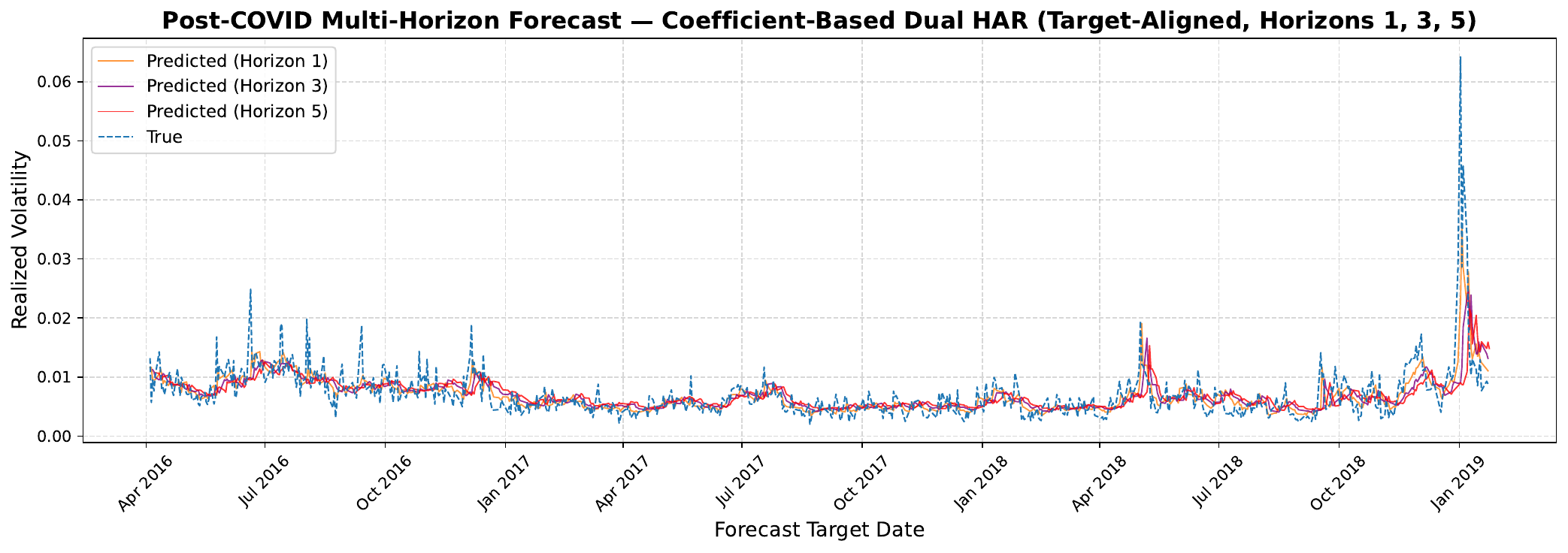}
        \caption{Post-COVID: Dual-Recursive Coefficient-Based Clustering.}
        \label{fig:post-5day}
    \end{subfigure}

    \caption{Forecasts from the optimally performing 5-day recursive forecasting model across all periods. Models correspond to results in Table \ref{tab:5-day-recursive}.}
    \label{fig:recursive-5day}
\end{figure*}


\begin{table}[H]
  \centering
  \begin{tabular}{lccc|ccc|ccc}
    \toprule
    \multicolumn{1}{c}{Model} & \multicolumn{3}{c}{Pre-COVID} & \multicolumn{3}{c}{COVID} & \multicolumn{3}{c}{Post-COVID}\\
    \midrule
    Single-Recursive & MAPE & MSE & \# Reg. & MAPE & MSE & \# Reg. & MAPE & MSE & \# Reg. \\
    \midrule
    Standard HAR        & 38.19 & 6.22 & 1 & 40.03 & 101.7 & 1 & 34.90 & 21.48 & 1 \\
    Markov Soft EM      & 38.18 & 6.22 & 2 & 42.98 & 96.96 & 2 & 34.85 & 20.94 & 2 \\
    Dist. Clustering    & 39.39 & 6.47 & 2 & 38.46 & 98.20 & 2 & 36.70 & 23.69 & 2 \\
    Coeff. Clustering   & 39.45 & 5.51 & 2 & 39.40 & 101.7 & 2 & 33.91 & 15.59 & 2 \\
    \midrule
    Dual-Recursive & MAPE & MSE & \# Reg. & MAPE & MSE & \# Reg. & MAPE & MSE & \# Reg. \\
    \midrule
    Standard HAR        & 37.83 & 6.17 & 1 & 52.39 & 155.0 & 1 & 32.25 & 16.44 & 1 \\
    Markov Soft EM      & 37.04 & 6.15 & 2 & 60.39 & 149.7 & 2 & 31.41 & 16.04 & 2 \\
    Dist. Clustering    & 34.41 & 6.19 & 2 & 47.79 & \cellcolor{yellow}77.85 & 2 & 32.21 & 18.30 & 2 \\
    Coeff. Clustering   & 38.10 & \cellcolor{yellow}5.50 & 2 & 55.14 & 391.1 & 2 & 32.02 & \cellcolor{yellow}14.38 & 2 \\
    \bottomrule
  \end{tabular}
  \caption{Model performance for 10-day forecasts during each time period. Divide MSE values by 1,000,000. Performance is compared based on MSE values.}
  \label{tab:10-day-recursive}
\end{table}

\begin{figure*}[!t]
    \centering

    \begin{subfigure}[t]{0.48\textwidth}
        \centering
        \includegraphics[width=\linewidth]{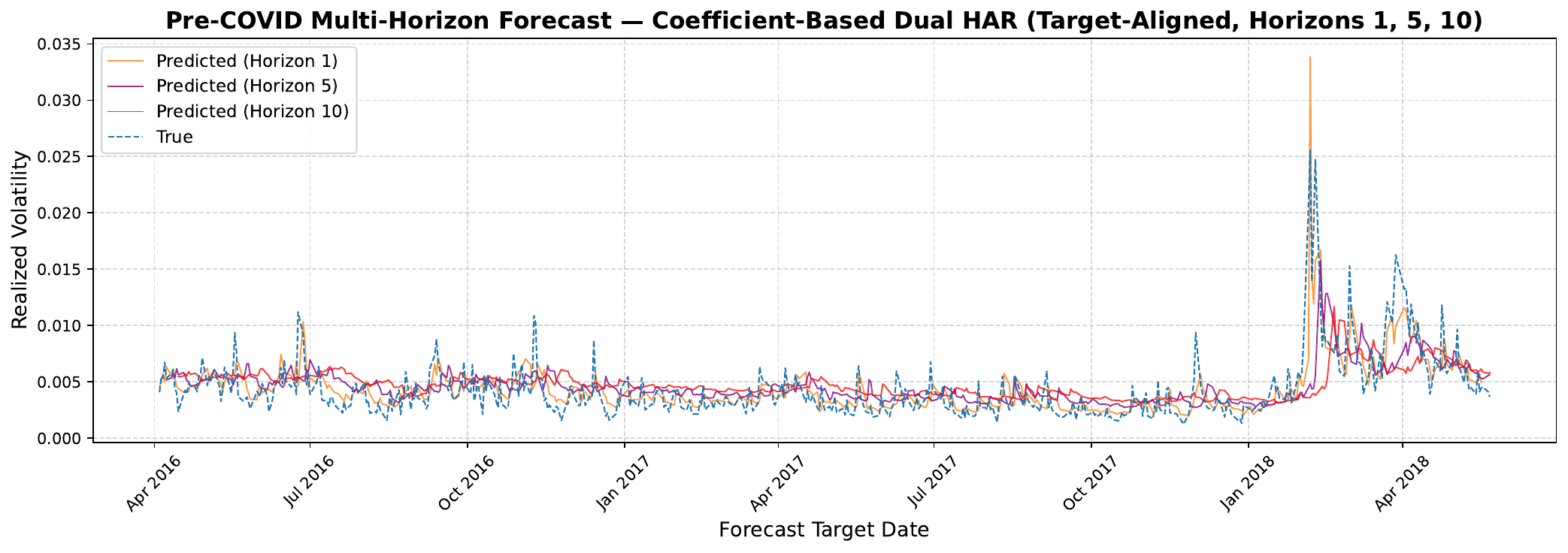}
        \caption{Pre-COVID: Dual-Recursive Coefficient-Based Clustering.}
        \label{fig:pre-10day}
    \end{subfigure}%
    \hfill
    \begin{subfigure}[t]{0.48\textwidth}
        \centering
        \includegraphics[width=\linewidth]{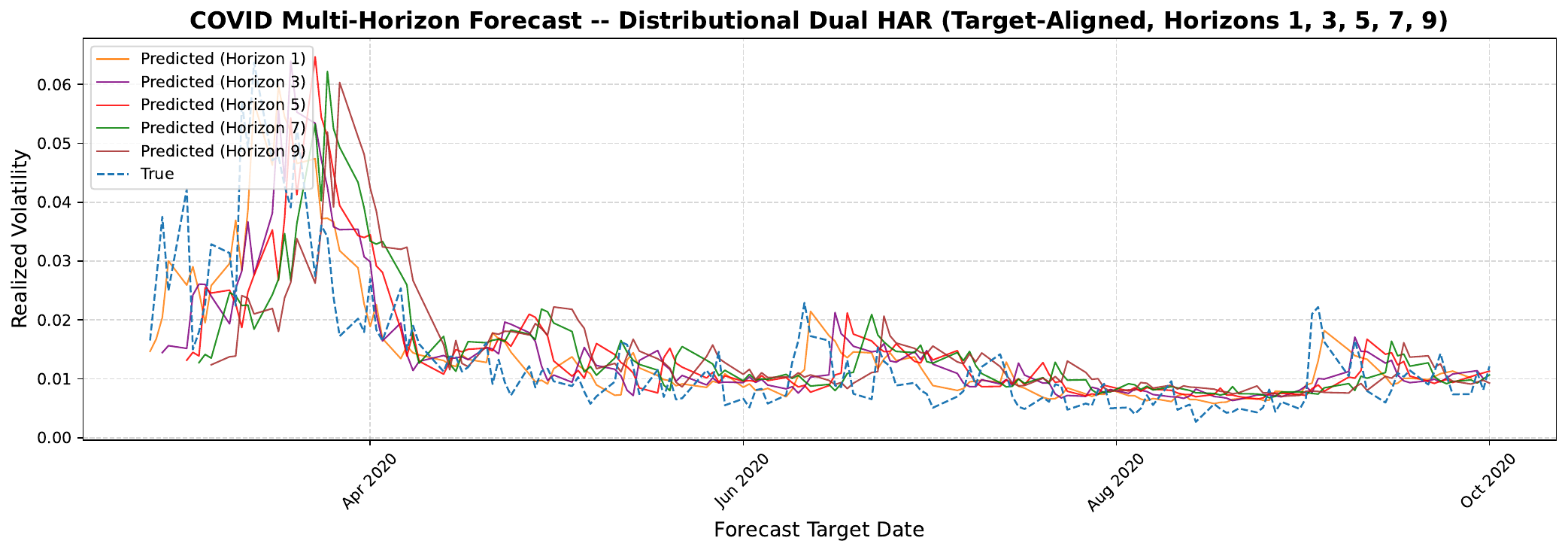}
        \caption{COVID: Dual-Recursive Distributional Clustering.}
        \label{fig:covid-10day}
    \end{subfigure}

    \vspace{1em}

    \begin{subfigure}[t]{0.6\textwidth}
        \centering
        \includegraphics[width=\linewidth]{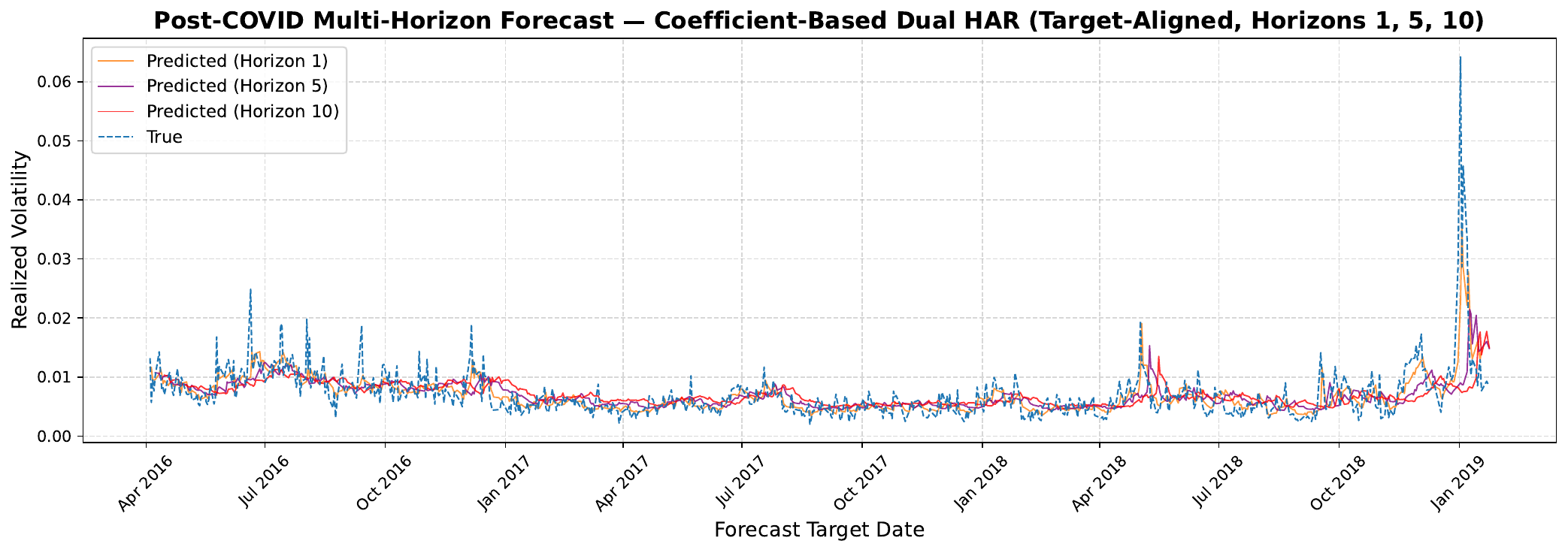}
        \caption{Post-COVID: Dual-Recursive Coefficient-Based Clustering.}
        \label{fig:post-10day}
    \end{subfigure}

    \caption{Forecasts from the optimally performing 10-day recursive forecasting model across all periods. Models correspond to results in Table \ref{tab:10-day-recursive}.}
    \label{fig:recursive-10day}
\end{figure*}

In the 5-day recursive forecasting models, visualized in Figure \ref{fig:recursive-5day}, the Coefficient-based Clustering model implemented with a dual-recursive structure achieves the lowest MSE of $4.46 \times 10^{-6}$ in the Pre-COVID period. This highlights the benefit of allowing regime-specific coefficients to reflect the stable relationships present during calm market conditions. During the COVID period, which is characterized by heightened volatility and structural breaks, the Distributional Clustering model with dual-recursive forecasting delivers the most accurate predictions. This result suggests that segmenting regimes based on distributional properties provides robustness in environments that are marked by sharp changes in volatility behavior. In the Post-COVID period, the Coefficient-based Clustering model again outperforms all others under the dual-recursive framework, with an MSE of $12.17\times 10^{-6}$, indicating that this model adapts well to normalized but structurally evolved post-crisis dynamics.

In the 10-day recursive setting seen in Figure \ref{fig:recursive-10day}, forecasting accuracy generally declines, as expected, due to the increased uncertainty associated with longer horizons. However, the patterns across models remain consistent. The Coefficient-based Clustering model under the dual-recursive structure again achieves the best performance in the Pre-COVID period. During the COVID period, the Distributional Clustering model performs best, mirroring its relative advantage under stress conditions observed in the 5-day forecasts. In the Post-COVID period, the Coefficient-based Clustering model once again leads.

Across both forecasting horizons, dual-recursive implementations consistently outperform their single-recursive counterparts. By allowing both series (realized volatility and implied volatility) to evolve together and by capturing interdependent dynamics over time, the dual-recursive framework provides a significant forecasting advantage. Notably, the Coefficient-based Clustering model performs best during the more stable Pre- and Post-COVID periods, while the Distributional Clustering model proves more resilient during the turbulent COVID period. Meanwhile, the Markov Soft EM model, while theoretically grounded, underperforms relative to clustering-based methods, especially during volatile periods. The superior performance of clustering-based models under a dual-recursive framework aligns with performances of the models in a non-recursive setting. These outcomes thus underscore the importance of flexible, adaptive structures that can effectively capture the nonlinear and regime-dependent properties of volatility.

\section{Limitations}
Several methodological limitations should be considered when interpreting the results of these regime-switching frameworks. Firstly, forecast accuracy is sensitive to hyperparameter choices, including the number of regimes, the window size for the Mood test, the dimensionality retained through PCA, and the tuning of the XGBoost classifier. Although these were guided by prior literature and cross-validation, alternate configurations may yield different results. Second, the projection of segment-level regime labels onto individual time points in clustering assumes behavioral homogeneity within each segment, which may oversimplify regime transitions near boundaries. However, this is used as a type of smoothing, and therefore the assumption of behavioral homogeneity also functions as a technique for mitigating microstructure noise. Another limitation to be considered is that in coefficient-based clustering, PCA is used to reduce the dimensionality of local OLS coefficients before clustering. While this helps avoid overfitting, it may obscure complex feature-volatility relationships critical to effective regime identification. Our Markov and distributional regime detection methods also rely heavily on realized volatility dynamics, which, while informative, may omit other structural indicators represented through the HAR features. This is potentially why coefficient-based clustering outperforms in  cases, as all three method's HAR models are fitting on the features, however, only the coefficient-based model clusters on feature information as well. Finally, the models exhibit limited robustness to extreme events. Although performance improves relative to baseline models during turbulent periods like COVID, methods such as the HMM may still lag in adapting to sudden structural breaks or nonlinear transitions, as reflected in reduced accuracy for the Markov Soft EM model.

\section{Conclusion}

Our empirical results demonstrate that regime-aware HAR extensions consistently yield lower forecasting errors than the standard HAR model, confirming the value of incorporating regime-switching dynamics. Specifically, coefficient-based soft clustering effectively captures gradual shifts and identifies structural breaks in volatility distributions, achieving superior performance before, during, and after the COVID time period. The Markov Soft EM model marginally outperforms the HAR model before and after the COVID time period, yet struggles to accurately capture volatility behavior during the highly volatile COVID-19 time period. Distributional clustering likewise demonstrates slight improvements against the HAR model. Inclusion of the VIX as a forward-looking feature was also shown to enhance model responsiveness to shifts in market sentiment, thereby refining predictive accuracy.

These outcomes illustrate that volatility dynamics exhibit regime-dependent statistical properties, and that flexible models including regime-specific parameters provide a meaningful advantage. In particular, clustering on the relationship between features and next-day volatility offers a novel and advantageous approach to regime-switching forecasting. Among the approaches evaluated, coefficient-based clustering stands out as the most effective method, due to its ability to adapt to nonstationarities in the data and maintain robustness across different market conditions. Its success suggests that modeling the evolving structure of the feature-volatility relationship—rather than static distributions—yields a more accurate and adaptive volatility forecasting framework.

These results have practical relevance for financial forecasting and risk management. The ability of coefficient-based clustering to adapt to structural breaks and changing feature-volatility relationships suggests potential value in applications such as short-term risk estimation. Moreover, the clustering-based regime-switching approach—especially in its coefficient-driven form—shows promise for generalization across different financial assets, as it relies on relational dynamics rather than asset-specific assumptions. Given the structural similarities in volatility processes across asset classes, extending this framework to equities, commodities, or cryptocurrencies could further validate its utility.

The regime-switching models show only marginal improvement over the feature-engineered HAR model, as it already performs fairly strongly. Future research could explore hybrid frameworks that integrate clustering and Markov switching, as well as advanced sequential models like LSTMs to capture more complex temporal dependencies and thereby see larger improvements.

\section{Acknowledgments}
This research was conducted as part of the Columbia Summer Undergraduate Research Experiences in Mathematical Modeling (CSUREMM) program, hosted and supported by the Columbia University Department of Mathematics. We would like to thank our mentors--Professor George Dragomir, Vihan Pandey, and Professor Dobrin Marchev--for their guidance, insight, and support throughout the course of this project.

\bibliography{refs}

\section*{Appendix}

\appendix

\section{Feature Description}
\label{appx:features}
\begin{table}[H]
\centering
\caption{\centering Model Feature Description and Calculation}
\label{tab:features}
\begin{threeparttable}
\begin{tabular}{>{\raggedright\arraybackslash}p{4cm} p{8cm} p{3cm}}
\hline
\textbf{Feature} & \textbf{Description} & \textbf{Calculation} \\
\hline
Daily Realized Volatility ($RV$) 
    & Measures the actual observed volatility of the S\&P 500 on a daily basis, computed from high-frequency intraday returns. Reflects short-term market risk and variability. 
    & See Eq. \ref{eq:RVt}. \\
Weekly Realized Volatility ($\overline{RV}^{(w)}$) 
    & Averages daily realized volatility over the past trading week to capture medium-term volatility trends and smooth short-term noise.
    & $\frac{1}{5} \sum_{i=0}^{4} RV_{t - i}$ \\
Monthly Realized Volatility ($\overline{RV}^{(m)}$) 
    & Averages daily realized volatility over the past trading month to identify longer-term volatility patterns and regime shifts.
    & $\frac{1}{22} \sum_{i=0}^{21} RV_{t - i}$ \\
Daily VIX ($VIX$) 
    & The CBOE Volatility Index value for the day, representing the market's expectation of near-term (30-day) volatility derived from option prices.
    & Sourced directly from VIX Index. \\
Weekly VIX ($\overline{VIX}^{(w)}$) 
    & The average of daily VIX values over the past trading week, smoothing out daily fluctuations to better capture medium-term market sentiment.
    & $\frac{1}{5} \sum_{i=0}^{4} VIX_{t - i}$ \\
Monthly VIX ($\overline{VIX}^{(m)}$) 
    & The average of daily VIX values over the past trading month, used to capture longer-term implied volatility trends and investor fear/uncertainty levels.
    & $\frac{1}{22} \sum_{i=0}^{21} VIX_{t - i}$ \\
Realized Kurtosis ($KTS$)
    & A statistical measure quantifying the ``tailedness'' or extremity of the return distribution, highlighting the presence of heavy tails or extreme events in returns. 
    & $\frac{N \sum_{i=1}^N r_{t,i}^4}{\left(\sum_{i=1}^N r_{t,i}^2\right)^2}$ \hspace{0.1cm}\tnote{$^*$} \\
Jump Variation ($JMP$) 
    & Quantifies discontinuous jumps in price returns separate from continuous volatility, capturing sudden market moves or shocks.
    & $RV_t - BV_t$ \hspace{0.1cm}\tnote{\textdagger} \\
\hline
\end{tabular}

\begin{tablenotes}
\footnotesize
\item[$^*$] \( r_{t,i} \) is the \( i \)-th intraday return on day \( t \), and \( N \) is the total number of intraday returns for day \( t \).
\item[\textdagger] $BV_t$ is bipower variation, calculated as $\frac{\pi}{2} \sum_{i=2}^N |r_{t,i}| \cdot |r_{t,i-1}|$
\end{tablenotes}
\end{threeparttable}
\end{table}

\section{XGBoost Hyperparameters}
\label{appx:hyperparams}
\begin{table}[H]
\centering
\caption{\centering Hyperparameters Used in XGBoost for Distributional\\and Coefficient-Based Clustering}
\label{tab:hyperparams}
\begin{tabular}{p{3.5cm} p{7.5cm} p{4cm}}
\hline
\textbf{Hyperparameter} & \textbf{Description} & \textbf{Search Space} \\
\hline
\texttt{n\_estimators} & Number of trees (boosting rounds) in the ensemble. More trees can improve performance but may increase overfitting and training time. & \{50, 100, 200, 300\} \\
\texttt{max\_depth} & Maximum depth of each decision tree. Controls model complexity; deeper trees can model more complex relationships but risk overfitting. & \{3, 5, 7, 10\} \\
\texttt{learning\_rate} & Also called eta; scales the contribution of each tree. Lower values slow down learning for more robust fitting, often requiring more trees. & \(\text{np.linspace}(0.01, 0.3, 10)\) \\
\texttt{subsample} & Fraction of training samples randomly selected for each boosting round. Helps prevent overfitting by introducing randomness. & \(\text{np.linspace}(0.6, 1.0, 5)\) \\
\texttt{colsample\_bytree} & Fraction of features (columns) randomly selected for each tree. Reduces correlation among trees and overfitting risk. & \(\text{np.linspace}(0.6, 1.0, 5)\) \\
\texttt{gamma} & Minimum loss reduction required to make a split. Larger values make the algorithm more conservative, pruning trees more aggressively. & \{0, 0.1, 0.2, 0.3\} \\
\texttt{reg\_alpha} & L1 regularization term on weights (Lasso). Encourages sparsity in feature weights, helping feature selection and reducing overfitting. & \{0, 0.01, 0.1, 1\} \\
\texttt{reg\_lambda} & L2 regularization term on weights (Ridge). Penalizes large weights, reducing model complexity and overfitting. & \{1, 1.5, 2, 3\} \\
\hline
\end{tabular}
\end{table}

\section{Libraries Used}
\label{appx:libraries}
\begin{table}[H]
\centering
\caption{\centering Python Libraries Used in the Project}
\label{tab:libraries}
\begin{tabular}{p{4cm}p{11cm}}
\hline
\textbf{Library} & \textbf{Description} \\
\hline
\texttt{matplotlib} & Used to visualize time series forecasts, model performance comparisons, and residual plots. \\
\texttt{NumPy} & Performs core numerical operations, array manipulation, and matrix mathematics. \\
\texttt{pandas} & Used for time series data manipulation, rolling window computations, and data handling via DataFrames. \\
\texttt{SciPy} & Enables statistical testing, including the Mood test for detecting distributional changes across segments. \\
\texttt{statsmodels} & Implements OLS regression and model diagnostics; used to fit and evaluate standard and regime-specific HAR models. \\
\texttt{scikit-learn} & Used for clustering, feature scaling, and model selection, including cross-validation and hyperparameter tuning. \\
\texttt{xgboost} & Implements gradient boosting framework; used for predicting soft regime probabilities. \\
\texttt{hmmlearn} & Implements Hidden Markov Models, used for inferring latent volatility regimes via Gaussian emissions. \\
\texttt{joblib} & Used to save and load Python objects such as scalers, models, and clustering outputs. \\
\texttt{POT} & Provides tools for optimal transport; used to compute Wasserstein distances between segment distributions. \\
\hline
\end{tabular}
\end{table}

\end{document}